\documentclass[conference]{IEEEtran}
\IEEEoverridecommandlockouts
\usepackage{cite}
\usepackage{amsmath,amssymb,amsfonts}
\usepackage{algorithmic}
\usepackage{graphicx}
\usepackage{textcomp}
\usepackage{xcolor}
\usepackage{lipsum}
\usepackage{multirow}
\usepackage{array} 
\usepackage{tabularx}
\usepackage{todonotes}
\usepackage{enumitem}
\usepackage{longtable}
\usepackage{tabularx, colortbl}
\usepackage{balance}

\def\BibTeX{{\rm B\kern-.05em{\sc i\kern-.025em b}\kern-.08em
    T\kern-.1667em\lower.7ex\hbox{E}\kern-.125emX}}
\begin{document}

\title{Leveraging Digital Twin Technologies for Public Space Protection and Vulnerability Assessment \thanks{The research leading to these results received funding from the European Commission under Grant Agreement No. 101073876 (Ceasefire).}}

\makeatletter
\newcommand{\linebreakand}{%
  \end{@IEEEauthorhalign}
  \hfill\mbox{}\par
  \mbox{}\hfill\begin{@IEEEauthorhalign}
}
\makeatother

\author{
  \IEEEauthorblockN{Artemis Stefanidou}
  \IEEEauthorblockA{\textit{Department of Informatics and Telematics~~}\\
    \textit{Harokopio University}\\
    Athens, Greece \\
    astefanidou@hua.gr}
  \and
  \IEEEauthorblockN{Jorgen Cani~~~~~~}
  \IEEEauthorblockA{\textit{Department of Informatics and Telematics~}\\
    \textit{Harokopio University~~~~~}\\
    Athens, Greece ~~~~~ \\
    cani@hua.gr ~~~~~}
  \and
  \IEEEauthorblockN{Thomas Papadopoulos~~}
  \IEEEauthorblockA{\textit{~~~~Regional Unit of Kastoria~~~~~} \\
    \textit{Region of Western Macedonia~~} \\
    Kastoria, Greece ~ \\
    thomaspapado@gmail.com~~}
  \linebreakand 
  \IEEEauthorblockN{~~~Panagiotis Radoglou-Grammatikis}
  \IEEEauthorblockA{\textit{~~~~~Department of Electrical and}\\
    \textit{~~~Computer Engineering} \\
    \textit{~~~University of Western Macedonia}\\
    ~~~Kozani, Greece \\
    ~~~pradoglou@uowm.gr}
  \and
  \IEEEauthorblockN{~~~~~~~~Panagiotis Sarigiannidis}
\IEEEauthorblockA{\textit{~~~~~~~~Department of Electrical and}\\
    \textit{~~~~~~~~~Computer Engineering} \\
    \textit{~~~~~~~~University of Western Macedonia}\\
    ~~~~~~~~Kozani, Greece \\
    ~~~~~~~~psarigiannidis@uowm.gr}
  \and
  \IEEEauthorblockN{~~~~~Iraklis Varlamis}
  \IEEEauthorblockA{\textit{~~Department of Informatics and Telematics} \\
    \textit{~~~~~~Harokopio University}\\
    ~~~~~~Athens, Greece \\
    ~~~~~~varlamis@hua.gr}
     \linebreakand 
    \IEEEauthorblockN{~~~~~~~~~~~~}
    \IEEEauthorblockA{~~~~~~~~~~~~~~}
    \and
    \IEEEauthorblockN{Georgios Th. Papadopoulos}
    \IEEEauthorblockA{\textit{Department of Informatics and Telematics} \\
    \textit{Harokopio University~~}\\
    Athens, Greece ~~\\
    g.th.papadopoulos@hua.gr~~}
    \and
    \IEEEauthorblockN{~~~~~~~~~~~}
    \IEEEauthorblockA{~~~~~~~}
}


\maketitle

\begin{abstract}
In recent years, the protection of so-called "soft targets", has become an increasingly important and challenging issue. The complexity and seriousness of this security threat have been growing exponentially, particularly with the advent of advanced technologies such as Artificial Intelligence (AI), Autonomous Vehicles (AVs), and 3D printing, especially in the context of large-scale, popular, and diverse public spaces. In this paper, a novel Digital Twin-as-a-Security-Service (DTaaSS) architecture is introduced for holistically and significantly enhancing the protection of public spaces (e.g. metro stations, leisure sites, urban squares, etc.). The proposed framework combines a Digital Twin (DT) conceptualization with additional cutting-edge technologies, including Internet of Things (IoT), cloud computing, Big Data analytics and AI. In particular, DTaaSS comprises a holistic, real-time, large-scale, comprehensive and data-driven security solution for the efficient/robust protection of public spaces, supporting: a) data collection and analytics, b) area monitoring/control and proactive threat detection, c) incident/attack prediction, and d) quantitative and data-driven vulnerability assessment. Overall, the designed architecture exhibits increased potential in  handling complex, hybrid and combined threats over large, critical and popular soft-targets. The applicability and robustness of DTaaSS is discussed in detail against representative and diverse real-world application scenarios, including complex attacks to: a) a metro station, b) a leisure site, and c) a cathedral square.
\end{abstract}


\section{Introduction}
Over the recent period, an increasingly rising number of attacks to various common and popular public spaces, due to different types of ideological, political, economic and social motivation, has been observed. To this end, the protection of the so-called `soft-targets', i.e. locations easily accessible by the general public with relatively low, though, security measures, has emerged as a rather challenging and increasingly important issue. To make the situation worse, the complexity and seriousness of this security threat further growths in an exponential way, due to the emergence of new advanced technologies (e.g. Artificial Intelligence (AI) \cite{rodis2023multimodal}, Autonomous Vehicles (AVs), 3-Dimensional (3D) printing, etc.); especially when it comes to large-scale, popular and diverse public spaces.

The current dominant paradigm and common practise for protecting `soft-targets' largely relies on the concept of `Security-by-Design' \cite{coaffee2022security}, introduced by the Joint Research Centre (JRC) of the European Commission (EC). In particular, this methodology is grounded on the fundamental principle of addressing security features from the very beginning of the conception and design of public locations, taking into account their inherent openness and integration to the overall urban landscape. Its main goal is to facilitate towards balancing efforts to increase urban resilience, whilst promoting the open and inclusive character of the public sphere. However, the `Security-by-Design' concept exhibits two major drawbacks: a) It relies only on the exploitation of domain-specific and empirical knowledge, i.e. ignoring current computational tools (e.g. simulation engines) that could provide valuable insights, and b) It does not take into account modern technological breakthroughs and innovations, which can be used by both security agencies and potential attackers.

\begin{table*}[t]
\centering
\small
\caption{Overview of key applications of DT in the security domain}
\label{table1}
\renewcommand{\arraystretch}{1.5} 
\setlength{\tabcolsep}{2pt} 
\begin{tabularx}{\linewidth}{|>{\centering\arraybackslash}m{1.2cm}|>{\arraybackslash}m{5.4cm}|>{\arraybackslash}m{4cm}|>{\arraybackslash}m{3cm}|>{\arraybackslash}m{3.75cm}|}
\hline
\rowcolor[HTML]{A9A9A9} 
\textbf{Method} & \centering \textbf{Task} & \centering \textbf{Model architecture} & \centering \textbf{Additional technologies} & \hspace{17pt} \textbf{Application field} \\
\hline
\rowcolor[HTML]{DCDCDC} 
\cite{gehrmann_digital_2020} & Data sharing and control of security-critical processes, state synchronization, protected software upgrade & DT replication model, security architecture & \centering Security components & Software upgrade, security critical processes \\
\hline
\rowcolor[HTML]{F0F0F0} 
\cite{danilczyk_angel_2019} & Microgrid security, real-time monitoring and control, diagnosis and mitigation of failures and cyber attacks & ANGEL DT framework, real-time, physics-based simulation & Microgrid technology, real-time data visualization & Microgrid security, cyber-physical system resilience \\
\hline
\rowcolor[HTML]{DCDCDC}
\cite{serral_practice_2021} & Cybersecurity in early development phases, cost-effective countermeasure analysis & Cybersecurity DT, enterprise architecture model & \centering Security analysis & Critical infrastructure security, intelligent transport systems \\
\hline
\rowcolor[HTML]{F0F0F0}
\cite{khajavi_digital_2023} & Fire and anomaly detection, building safety and security, data analysis & DT for safety and security, 3D visualizations and sensors & Advanced sensors, data analysis, 3D visualization & Building safety, fire and anomaly detection \\
\hline
\rowcolor[HTML]{DCDCDC}
\cite{sousa_elegant_2021} & Off-premises DTs, data collection and processing, security threat mitigation & Cloud-based DT architecture, high fidelity replicas, data pipelines & Cloud services, machine learning & Critical infrastructure security, denial of service attacks \\
\hline
\rowcolor[HTML]{F0F0F0}
\cite{becue_cyberfactory1_2018} & Balancing productivity and security, simulation and optimization, resilience improvement & DT integration with cyber-range & Cyber-range simulation & Manufacturing systems, avionics, robotic systems \\
\hline
\rowcolor[HTML]{DCDCDC}
\cite{empl_digital-twin-based_2023} & Alignment of security analytics with DTs, shareable cybersecurity knowledge & DT2SA model, formal model for security analytics & 
 \centering Security analytics & IoT security, cybersecurity knowledge sharing \\
\hline
\rowcolor[HTML]{F0F0F0}
\cite{dietz_integrating_2020} & Integration of DT simulations in enterprise security, attack simulation, SOC support & Process-based security framework, DT security simulation & \centering SOC integration & Industrial asset monitoring, attack simulation \\
\hline
\rowcolor[HTML]{DCDCDC}
\cite{oconnell_digital_2023} & Interoperability and security, seamless system of systems & Emerging DT standards, system of systems architecture & \centering Industry 4.0 networks & Smart manufacturing systems, industry 4.0 networks \\
\hline
\rowcolor[HTML]{F0F0F0}
\cite{fraser_enhancing_2021} & Real-time intrusion and anomaly detection, intrusion detection validation & DT architecture, machine learning models & \centering Data-driven methods & UAV security, anomaly detection \\
\hline
\rowcolor[HTML]{DCDCDC}
\cite{eckhart_towards_2018} & Security and safety rules monitoring, testing and exploring systems & DT framework, automated generation, security modules & Data exchange formats & Cyber-physical systems, security and safety assessment \\
\hline
\end{tabularx}
\end{table*}

Besides purely security measure enforcement, the tremendous recent technological advancements driven by the `Industry 4.0' realm have dramatically transformed the way that management and monitoring of physical systems is being performed. Specifically, innovations in fields like sensor-based analysis, cloud computing and data analytics have enabled the creation of virtual representations of physical assets; hence, facilitating real-time data collection/analysis and remote interaction/monitoring of physical systems \cite{tao2017digital,bruynseels2018digital,alazab2022digital}. The most widely adopted type of such virtual representations comprises the concept of `Digital Twins' (DTs) \cite{van2020taxonomy,tao2019digital}, which allows in-depth understanding and modeling of system behaviors and interactions. Despite the fact that DTs have been investigated/evaluated under different scenarios in the broader security application domain (e.g. data sharing and control \cite{gehrmann_digital_2020}, micro-grid security \cite{danilczyk_angel_2019}, cybersecurity \cite{serral_practice_2021}, etc.), they have not been used so far for enhancing the protection of public spaces.

In this paper, a novel Digital Twin-as-a-Security-Service (DTaaSS) architecture is designed for holistically and significantly enhancing the current security measures of `soft target' locations. DTaaSS conceptually extends the `Security-by-Design' paradigm, by incorporating modern, computational and AI-boosted technologies in the public space design, management and protection procedures. In particular, the proposed framework combines a DT formulation with additional cutting-edge technologies, including Internet of Things (IoT), cloud computing, Big Data analytics and AI. More specifically, DTaaSS comprises a holistic, real-time, large-scale, comprehensive and data-driven security solution for the robust protection of public spaces, exhibiting the following key advantageous characteristics:
\begin{itemize}
\item \textbf{Data collection and analytics}: It incorporates information from different types of sources, including a) physical devices and sensors (e.g. Sound navigation and ranging (Sonar), Closed Circuit TeleVision (CCTV), Radio Frequency (RF), biometric, mobile, etc.), b) cyber sources (e.g. social media, (dark) Web, forums, etc.), and c) security stakeholders' feedback (e.g. intelligence officers, patrol units, etc.).
\item \textbf{Area monitoring/control and proactive threat detection}: It includes AI-empowered tools for addressing a) physical world (e.g. Unmanned Aerial Vehicle (UAV) detection, mobile analytics, facial recognition, suspicious object detection, etc.) and b) cyber world (e.g. cyber patrolling, Natural Language Processing (NLP), cyber threat detection, etc.) activities.
\item \textbf{Incident/attack prediction}: Apart from timely threat detection, support is provided for identifying early signs and predictions of potential attacks.

\item \textbf{Quantitative and data-driven vulnerability assessment}: Complementary to the empirical `Security-by-design' approach, a simulation-based methodology (based on the DT scenario emulation capabilities) enables detailed and data-driven vulnerability assessment. 
\end{itemize}

\begin{figure}[t]
\centerline{\includegraphics[width=.9\linewidth]{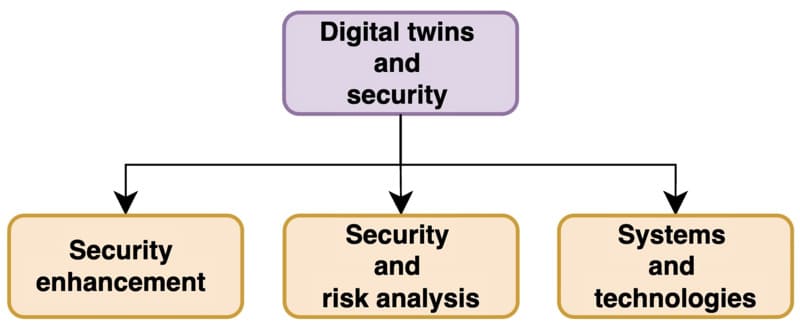}}
\caption{Categorization of DT applications in the security domain}
\label{diagramCateg}
\end{figure}

Overall, the designed architecture exhibits increased potential in  handling complex, hybrid and combined threats over large, critical and popular soft-targets. The applicability and robustness of DTaaSS is discussed in detail against representative and diverse real-world application scenarios, including the following types of complex attacks: a) bomb attack at metro station, b) hybrid Unmanned Surface Vehicles (USV) / Unmanned Underwater Vehicles (UUV) and shooting attack at leisure site, and c) combined UAV and suicide attack during celebration ceremony at cathedral square.

The remainder of the paper is organized as follows: Section \ref{sec:relwork} provides an overview of DT-based applications in the security domain. Additionally, Section \ref{sec:framework} details the proposed DTaaSS architecture, while Section \ref{sec:technologies} discusses its functional operation and individual components. Section \ref{sec:usecases} discusses the application of DTaaSS in representative real-world scenarios. Finally, Section \ref{sec:conclusions} concludes the paper and outlines future research directions.

\section{DT in security}
\label{sec:relwork}
The usage of DTs in the security domain has seen significant advancements over the recent years, enhancing system performance, safety and operational efficiency across various applications. These developments can be roughly categorized into three key areas, as illustrated in Fig. \ref{diagramCateg}:

\begin{itemize}
        \item \underline{Security enhancement}: Advancements in DTs that focus on improving space monitoring, early threat detection and designing control mechanisms to bolster overall safety.
        \item \underline{Security and risk analysis}: Use of DTs for evaluating and simulating security threats and vulnerabilities, leading to enhanced risk management and threat prediction/mitigation.
        \item \underline{Systems and technologies}: Integration of DTs within technological environments to enhance interoperability, real-time threat detection and cyber-security measures.
\end{itemize}

\subsection{Security enhancement}
Recent advancements in DT technology have significantly impacted security across various critical infrastructures. Gehrmann et al. \cite{gehrmann_digital_2020} present a security architecture specifically designed for digital twin systems, focusing on how replication models can enhance data sharing and critical process control. This approach highlights the role of DTs in supporting secure data exchange and controlled environments. In another study, the so-called Automatic Network Guardian for ELectrical systems (ANGEL) DT framework \cite{danilczyk_angel_2019} improves the security of microgrids by combining real-time simulations with physical systems. This integration enables continuous monitoring and control, underscoring the potential of DTs to enhance security resilience in electrical systems. Similarly, the application of DTs for building safety and security emphasizes their utility in fire detection and anomaly identification through advanced sensors and data analysis. The work of Khajavi et al. \cite{khajavi_digital_2023} illustrates how DTs can predict and manage potential safety threats throughout a building's lifecycle. Furthermore, Sousa et al. \cite{sousa_elegant_2021} extend these concepts by presenting a cloud-based approach for deploying DTs for critical infrastructure security. This approach demonstrates how scalable and efficient data collection processes support high-fidelity virtual replicas of physical systems, reducing costs and enhancing security through real-time monitoring and simulation.

\subsection{Security and risk analysis}
The integration of DT technology into modern infrastructure security strategies demonstrates its significant potential to bolster security and resilience of critical systems. The complex challenge of balancing productivity with security, by combining DTs with cyber-ranges, is discussed in \cite{becue_cyberfactory1_2018}. This approach showcases how DTs can be leveraged to enhance both resilience and security in manufacturing environments, ensuring that these systems can withstand and recover from potential threats. Additionally, a so called DT2SA model is introduced regarding IoT security in \cite{empl_digital-twin-based_2023}, which aligns DTs with advanced security analytics. Moreover, Dietz et al. \cite{dietz_integrating_2020} explore the application of DTs for monitoring and protection of industrial systems. The latter work proposes a framework that integrates security simulations with Security Operations Center (SOC) strategies, aiming to enhance defense against attacks and to better analyze their impacts on industrial operations.

\begin{figure*}[t]
\centerline{\includegraphics[width=.85\linewidth]{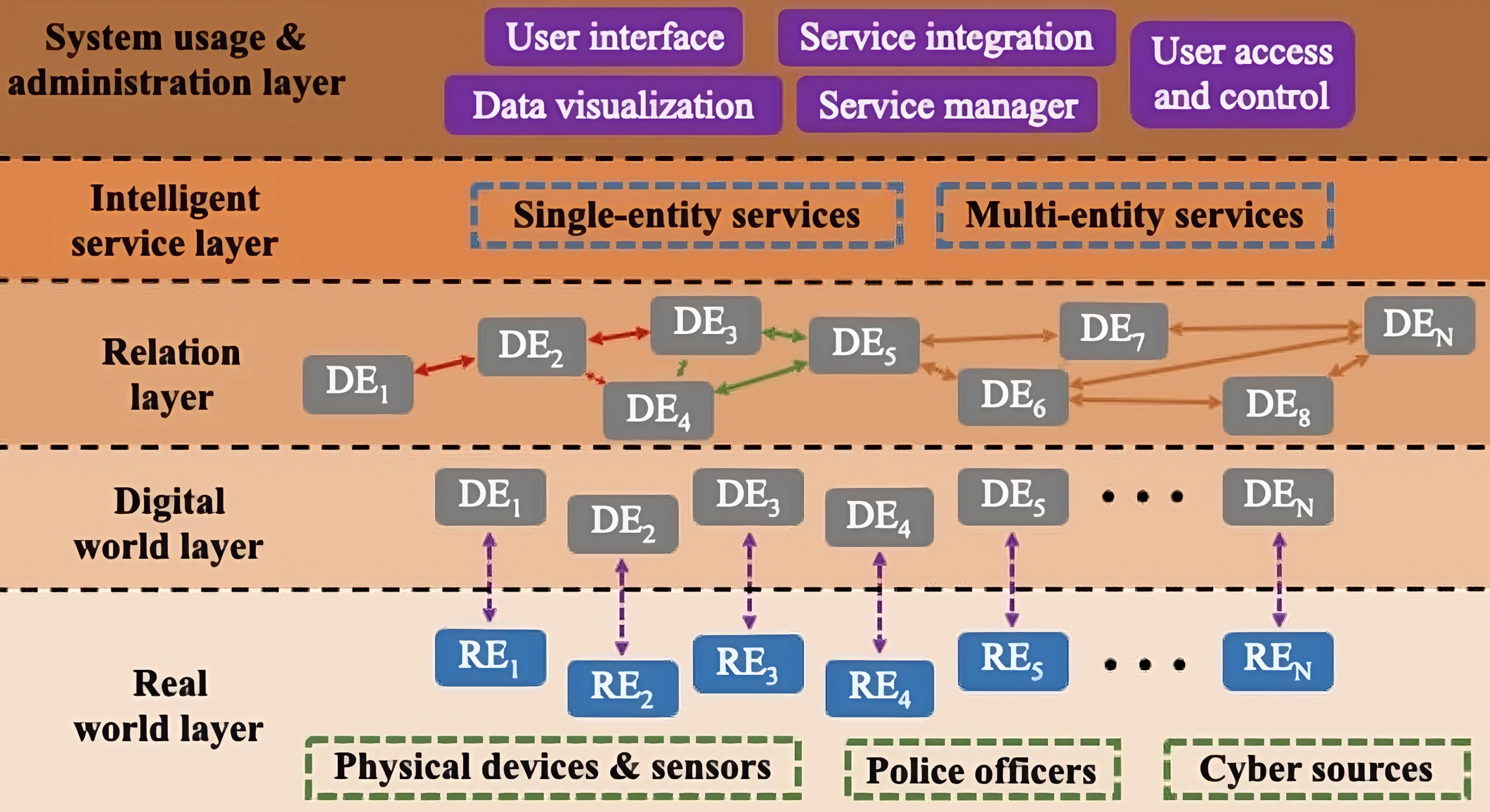}}
\caption{Proposed DTaaSS architecture for public space protection}
\label{arch}
\end{figure*}

\subsection{Systems and technologies}
The application of DTs in the realm of industrial systems and cyber threat management reveals their significant impact on enhancing security and resilience in complex environments. DTs facilitate the creation of interconnected systems through emerging standards, which fosters improved interoperability, resilience and security within Industry 4.0 networks. This integration addresses the need for cohesive and secure industrial operations \cite{oconnell_digital_2023}. Additionally, DTs are employed in the context of UAVs usage, where machine-learning techniques are used to perform real-time environment perception and anomaly detection. This application enhances the ability to detect and respond to potential security threats posed by UAVs \cite{fraser_enhancing_2021}. Moreover, the concept of a cybersecurity DT is proposed to simulate and analyze potential attacks on critical control systems in \cite{serral_practice_2021}. This approach allows for the development and testing of security measures in a virtual environment; thus, avoiding disruptions to physical infrastructure. Furthermore, a framework for creating and executing DTs of Cyber-Physical Systems (CPSs) is introduced in \cite{eckhart_towards_2018}. This framework automates the generation of virtual replicas from system specifications, using standardized engineering data exchange formats, while it enables security professionals to explore and to test simulated environments (without impacting live systems) and it includes integrated security modules to monitor CPS states and enforce security/safety protocols. 

A summarized overview of key DT-based approaches in the security domain is presented in Table \ref{table1}. Despite the large body of research across various applications, the use of DT technologies for the protection of public spaces, especially in the context of complex and hybrid attacks, has not been explored so far.

    \begin{figure}[h]
        \centerline{\includegraphics[width=.9\linewidth]{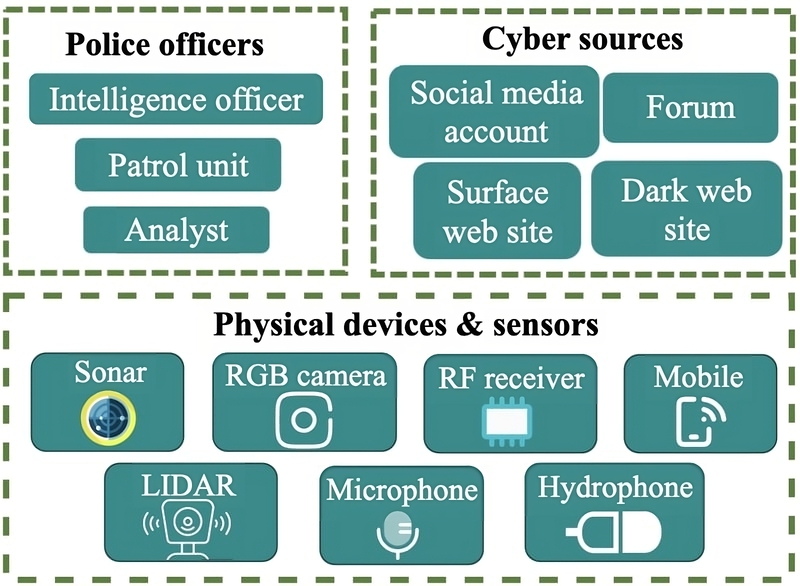}}
        \caption{Primary groups of entities in the real world layer}
        \label{realworld}
    \end{figure}
    
\section{DTaaSS architecture}
\label{sec:framework}

The proposed cloud-based DT architecture follows the IoT system design paradigm and allows the real-time collection, processing and analysis of data from a multitude of heterogeneous sources, in order to enhance public space security and protection. The architecture, graphically illustrated in Fig. \ref{arch}, consists of five interconnected and closely interacting layers, each playing a crucial role in the system:

\begin{enumerate}
    \item \textbf{Real world layer}: Like `things' in the IoT design paradigm, all entities of relevance for public space protection are included in the real world layer. These entities (denoted $RE_n$ in Fig. \ref{arch}), can be divided into three main groups, as detailed in Fig. \ref{realworld}:

    \begin{itemize}
        \item \underline{Police officers}: Law enforcement officers, including patrol units, analysts and intelligence officers, who interact directly with the system by providing input or evidence, such as identifying suspicious events or capturing images of potential suspects.
        \item \underline{Cyber sources}: This includes data from social media accounts, surface websites, forums and dark web sites. These sources provide critical information related to the security of the examined public space that is circulated on online/cyber communication streams.
        \item \underline{Physical devices and sensors}: This includes a variety of devices such as Sonar, Light Detection And Ranging (LiDAR), microphones, RGB cameras and RF receivers. These devices continuously provide real-time sensor data from the environment, enhancing situational awareness.
    \end{itemize}

    \item \textbf{Digital world layer}: This layer complements the real-world layer, by providing a digital counterpart for each real-world entity. These digital entities (denoted $DE_n$ in Fig. \ref{arch}) meticulously model the physical attributes and functionalities of their real-world counterparts. This digital representation enables seamless transitions between the real and the digital worlds, facilitating advanced analysis, processing, simulation, visualization and manipulation of the entities and their properties.

    \item \textbf{Relation layer}: This layer plays a pivotal role by defining the intricate and heterogeneous relationships between entities, effectively modeling the interactions between real-world objects. These relationships can be spatial, temporal, semantic and additional ones, connecting digital entities across various categories. Interactions modeled in the digital world are mirrored in the real one and vice versa, ensuring synchronization and responsiveness to changes and actions.
    
    \item \textbf{Intelligent service layer}: The intelligent service layer comprises AI-empowered information processing pipelines. These pipelines analyze information from real-world entities to detect and to respond to various stimuli and potential threats. The functionalities supported by this layer are categorized, as depicted in Fig. \ref{intelligent} into:
    \begin{itemize}
        \item \underline{Single-entity services}: These services process information from individual digital entities. Examples include source analysis, visual analytics, natural language processing, open space analytics, UAV detection, USV/UUV detection, mobile analytics, facial biometrics and enclosed space analytics.
        \item \underline{Multi-entity services}: These services combine information from multiple digital entities to provide comprehensive outputs. Examples include cyber threat detection, area monitoring, event detection, attack prediction, vulnerability assessment and information sharing.
    \end{itemize}

    \begin{figure}[t]
    \centerline{\includegraphics[width=.65\linewidth]{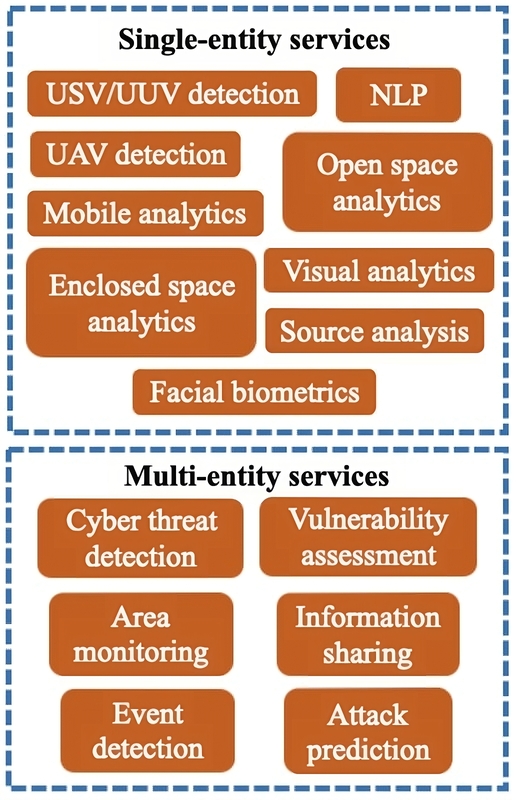}}
    \caption{Groups of functionalities supported by the intelligent service layer}
    \label{intelligent}
    \end{figure}

\begin{figure*}[htbp]
\centerline{\includegraphics[width=1\linewidth]{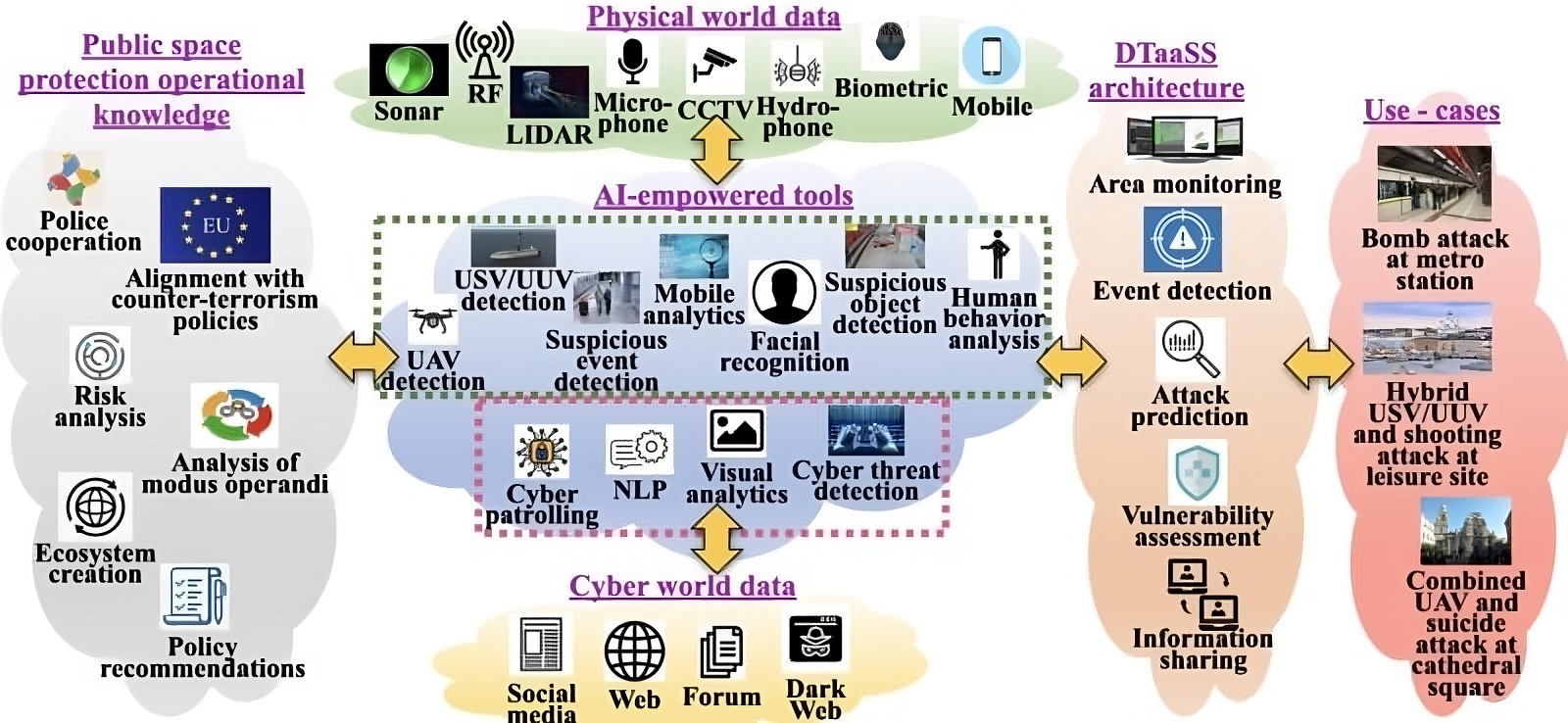}}
\caption{Graphical illustration of functional operations and individual components of the proposed framework}
\label{fig:functional}
\end{figure*}
    
    \item \textbf{System usage and administration layer}: This layer focuses on the practical application and management of the DTaaSS solution by end-users. It addresses aspects such as access control, operation, and integration with existing law enforcement systems and infrastructures. Key functionalities include user interfaces, data visualization, service integration, service management and user access control.
\end{enumerate}

\section{Functional operation and individual components}
\label{sec:technologies}

The functional architecture of the proposed framework, as depicted in Fig. \ref{fig:functional}, represents a comprehensive system designed to address security challenges across both the physical and the cyber domains. The system operation is structured so as to follow a logical flow, beginning with the collection and integration of diverse data sources, which form the foundation for the system's advanced AI-powered tools. These tools process the collected data to generate actionable insights, which are then provided as input to the DTaaSS architecture. The latter supports critical functionalities, such as area monitoring, event detection, attack prediction, vulnerability assessment and information sharing. It needs to be highlighted that the system also incorporates operational knowledge, ensuring that the technological capabilities are effectively applied in real-world scenarios. In the followings, each constituent component will be described in detail, outline how their integration facilitates towards forming a robust and adaptable security solution.

\subsection{Data sources}

The foundation of the system lies in the diverse data sources that are crucial for the effective functioning of the AI tools. These data sources are categorized into two main types, namely physical and cyber world data.

\underline{Physical world data}: This includes Sonar signals, which are sound propagation measurements collected at waterfront locations, RF receiver signals that refer to radio wave data used for aerial inspection and detection, and LiDAR capturings involving pulsed laser measurements that assist in effective object identification. Additional data types can be microphone recordings, comprising audio signals gathered from sensors in public spaces, and hydrophone recordings, containing audio data collected from surface and underwater sensors. Moreover, CCTV offers video surveillance data to detect abnormal or suspicious incidents and facial biometric traits for providing detailed data on facial characteristics for suspect identification.

\underline{Cyber world data}: This encompasses mobile network activity, including network data related to mobile location and data exchanges, social media information that covers data disseminated through social media platforms, and surface web information related to suspicious communication activities. It also involves forums discussing criminal-oriented topics and dark web data from the unindexed parts of the web associated with criminal-related activities. The integration of these data sources is crucial as they provide the necessary input for the system's AI tools, enabling the detection and analysis of potential threats in real time.

\subsection{AI-empowered tools}

Building on the rich data collected from the physical and the cyber worlds, the system incorporates a suite of AI-powered tools. These tools are specifically designed to process, analyze and derive actionable insights from the diverse data inputs, enabling the system to effectively address various security scenarios.

Regarding the processing of physical world data, the UAV detection tool facilitates aerial inspection and monitoring of UAVs, while USV/UUV detection focuses on identifying threats in waterfront areas, by detecting USVs and UUVs. To address potential threats in enclosed spaces, suspicious enclosed space event detection algorithms analyze activities within buildings or confined environments. Mobile data analytics provide valuable insights into real-world entities and incidents, utilizing network activities such as International Mobile Subscriber Identity (IMSI), subscriber details, base station registrations, and mobile device trajectories. These insights help recognize patterns and track the movement of individuals involved in incidents. At the same time, facial recognition technology uses biometric traits to accurately identify individuals by comparing them to a predefined list of suspects. Additionally, suspicious object detection applies advanced algorithms to identify potential threats, such as abandoned or illegal objects, using sensors like CCTV cameras. Finally, the open space human and crowd behavior analysis tool monitors the behavior of individuals and crowds in public spaces, detecting unusual activities and enabling timely responses for effective crowd management and the prevention of potential risks.

Concerning the analysis of cyber world data, cyber patrolling involves monitoring and collecting data from various online sources, including open-source and dark web platforms, forums and social media, in order to detect and analyze potential cyber threats \cite{miniadou2023investigating, mademlis2023invisible}. Additionally, NLP delves into online textual sources to extract meaningful insights \cite{tsirmpas2024neural,alexandridis2021survey}, while Big Data visual analytics examine extensive volumes of visual content in cyberspace to identify and assess threats \cite{konstantakos2023self,konstantakos2024self}. Moreover, cyber threat detection provides early warnings about potential cyber threats, by continuously monitoring online activities.

The above-described AI tools constitute the core processing units that convert raw data into meaningful intelligence, driving the system's ability to respond and to mitigate threats effectively.

\subsection{DTaaSS architecture}

In order to integrate the various AI technologies and to ensure seamless operation across different security scenarios, the overall framework relies on the DTaaSS architecture detailed in Section \ref{sec:framework}. In particular, DTaaSS is designed to support large-scale information fusion and the implementation of comprehensive procedures for public space protection, enabling the following key high-level functionalities:

\begin{itemize}
    \item \underline{Area monitoring}: This fuses data from multiple sources to provide a real-time and comprehensive intelligence picture of public spaces.
    \item \underline{Event detection}: This enables timely identification of suspicious activities, allowing for prompt countermeasures.
    \item \underline{Attack prediction}: This detects indicators of potential future attacks, based on observed patterns in cyber activities and historical data.
    \item \underline{Vulnerability assessment}: This makes use of the DT simulation capabilities to analyze vulnerabilities within the monitored environment.
    \item \underline{Information sharing}: This facilitates efficient information exchange among various police units and authorities, including links with national and international security systems.
\end{itemize}
Overall, the DTaaSS architecture not only consolidates various AI tools, but also provides the necessary infrastructure to ensure that data-driven insights are actionable and can be effectively communicated across different stakeholders. Additionally, the adopted software `as-a-service' design, allows the proposed framework's efficient use and easy integration to already available security operational systems.

\subsection{Public space protection operational knowledge}

In order to ensure the effective deployment of the DTaaSS architecture in real-world scenarios, it is essential to incorporate essential operational knowledge through a broad ecosystem of stakeholders. In particular, enhanced police cooperation is crucial for harmonizing procedures and aligning operational practices among various police authorities involved in public space protection. Additionally, alignment with counter-terrorism policies ensures that the system continuously adapts to relevant strategies and directives. In parallel, conducting a comprehensive risk analysis is necessary for systematically evaluating potential risks associated with public space protection, while an analysis of the criminal modus operandi can provide a deeper understanding of criminal phenomena and their connections to other forms of crime. Moreover, ecosystem creation involves establishing a broad network of stakeholders engaged in public space protection and counter-terrorism efforts. Furthermore, policy recommendations are essential for formulating suggestions to enhance police cooperation strategies and operational practices. Overall, operational knowledge constitutes a fundamental practical application component that ensures that the technological capabilities of the system are effectively utilized on the field.

\section{Representative real-world application scenarios}
\label{sec:usecases}

The proposed DTaaSS architecture, as detailed in Sections \ref{sec:framework} and \ref{sec:technologies}, is designed so as to be generic and highly-adaptable, while also being capable of handling various types of complex and hybrid security threats in public spaces. In this section, the applicability and robustness of DTaaSS is discussed in detail against representative and diverse real-world application scenarios, including the following types of complex attacks: a) bomb attack at metro station, b) hybrid USV/UUV and shooting attack at leisure site, and c) combined UAV and suicide attack during celebration ceremony at cathedral square.

\subsection{Use-case \#1: Bomb attack at metro station}

\underline{Aim}: The main objective is to develop and to implement effective methods for the timely detection of bomb attacks and the identification of the suspects involved. This requires the development of systems that can quickly recognize potential threats and accurately determine the identity of individuals that may be involved.

\underline{Motivation}: Public transportation stations/locations are typically characterized by high passenger density, especially during peak travel hours. This crowd density, combined with the rapid movement of commuters, complicates security monitoring efforts. Additionally, the dynamic and congested nature of such spaces makes it difficult to effectively observe and to respond to potential security threats, underscoring the need for advanced monitoring and detection measures.

\underline{Scenario}: Attacks to public transportation infrastructure (e.g. metro stations) constitute attractive criminal goals, due to their potential for causing a high number of casualties, especially during peak hours, and their relative easy accessibility. These attacks typically require significant planning and coordination. Perpetrators often conduct multiple reconnaissance visits to the target location over multiple weeks, gathering detailed information about security measures, attack routes and transportation schedules, as well as visual and multimedia data to aid their planning. During the attack's preparation phase, criminals/terrorists may position themselves near the target, waiting for the right moment and engaging in seemingly normal but actually preparatory behaviors, such as standing in specific spots over unusually long periods. They often use mobile phones for communication, including calls and information exchange with accomplices. When the attack occurs, multiple suspects enter the metro station following predetermined routes, attempting to blend in with the crowd, while carrying suspicious objects (e.g. large bags containing explosives). Despite their efforts to remain inconspicuous, their actions and appearances are captured by the station's CCTV cameras, which also monitor for abandoned items that could pose a threat.

\underline{Major challenges}: The following list of challenges, corresponding to functionalities/modules supported by the DTaaSS architecture, needs to be addressed for tacking such a criminal attack:

\begin{itemize}
    \item Detection of abnormal mobile communication patterns: Identification of unusual patterns in mobile communication data, such as location tracking and application usage statistics, can provide valuable insights to indicate suspicious behavior. This involves analyzing communication data to spot anomalies that could suggest the presence of individuals involved in planning or executing an attack.
    \item Facial recognition of suspects: Accurate identification of suspects through facial recognition technology constitutes essentially a critial aspect. This requires comparing images captured by CCTV cameras with a predefined database of known suspects to determine if any of them is present in the area.
    \item Detection of suspicious objects: Identifying abandoned or potentially dangerous objects within a metro station is crucial for preventing attacks. This involves using CCTV footage and other sensors to detect items left unattended by suspects.
    \item Identification of abnormal human behavior: Recognizing unusual or suspicious behavior, such as individuals standing still for extended periods of time or engaging in irregular movements, is essential for early threat detection. This requires analyzing patterns of behavior to identify actions that deviate from the norm.
    \item Handling high passenger density: Managing the high density of passengers presents a critical problem, as it can lead to significant occlusions in CCTV footage. This makes it difficult to clearly observe and identify individuals and objects in crowded environments. Addressing this issue involves developing techniques to improve the clarity and effectiveness of surveillance services in dense crowds.
\end{itemize}

\subsection{Use-case \#2: Hybrid USV/UUV and shooting attack at leisure site}

\underline{Aim}: The main goal is to develop methods for the early identification of abnormal activities, involving USVs and UUVs, as well as groups of suspects. This requires the development of systems that can effectively detect and differentiate between potentially threatening vehicles and ordinary vessels, as well as identifying suspicious human behaviors and objects.

\underline{Motivation}: Open-space leisure sites near sea-front locations are susceptible to attacks from multiple/different origins, including from the waterside.

\underline{Scenario}: Attacks to popular and highly-crowded touristic/leisure sites near coastlines during peak season, where the potential for high casualties and media impact is significant, constitutes a significant potential goal for criminals. Advances in 3D-printing technology have made it easier and affordable to construct homemade USVs and UUVs capable of carrying explosives. Detecting such unmanned vehicles amidst heavy maritime traffic and differentiating them from benign vessels or marine life is challenging. Simultaneously, land-based attacks using 3D-printed firearms may be carried out to increase casualties. Effective detection involves identifying unusual group behaviors and suspicious objects (e.g. large bags containing weapons), while also monitoring online forums and social media for early signs of attack planning.

\underline{Major challenges}: The following list of challenges, corresponding to functionalities/modules supported by the DTaaSS architecture, needs to be addressed for tacking such a criminal attack:

\begin{itemize}
    \item Early identification of suspicious USV/UUV activity: Detecting unusual activities involving USVs or UUVs, particularly in busy coastal areas where many vessels are present, and distinguishing them from other similar-looking vessels constitutes a difficult task.
    \item Identification of abnormal or suspicious human behavior: Recognizing unusual or suspicious behaviors of individuals or groups, such as repetitive or misleading actions and area inspections, can provide early warnings of a possible attack.
    \item Detection of suspicious objects: Identifying potentially dangerous objects, such as large packages or bags containing firearms and ammunition, at leisure sites constitutes a critical proactive measure.
    \item Encountering the presence of a large number of vessels: Managing the inherent challenge of significant visual occlusions and appearance similarities among numerous vessels in coastal areas may have a great operational impact.
    \item Detection of early signs of attack preparation activities on cyber channels: Monitoring online forums and social media for early indicators of attack preparation, including discussions about 3D-printed USVs/UUVs and firearms, can be proven valuable for encountering possible subsequent attacks.
\end{itemize}

\subsection{Use-case \#3: Combined UAV and suicide attack during celebration ceremony at cathedral square}

\underline{Aim}: The primary goal is to develop effective methods for the timely detection of abnormal activities, involving UAVs and individual suspects in crowded open-space religious sites.

\underline{Motivation}: Open-space religious sites, particularly during crowded celebration ceremonies, present significant vulnerabilities to terrorist attacks. These events, often involving large gatherings of people, provide a strategic opportunity for terrorists to inflict substantial harm and to spread their messages. The high density of participants and the extensive area covered during such ceremonies make them susceptible to attacks from various origins and through different means.

\underline{Scenario}: Celebration ceremonies, such as annual city-saint festivities, provide terrorists with opportunities to maximize casualties and to advance their agendas. UAVs, due to their versatility and low cost, can be exploited for malicious purposes, beyond their typical use for aerial footage capturing. They pose a severe threat by potentially transporting explosives discreetly into crowded areas, such as cathedral squares. Distinguishing between malicious and benign UAVs during highly-publicized events with extensive live TeleVision (TV) coverage is challenging. Additionally, lone wolf attackers armed with high-powered explosives may coordinate their actions to amplify the impact of their assault. Identifying these individuals amidst a dense crowd is particularly difficult. In parallel, early detection of suspicious online behavior, including forum discussions and social media activity related to attack preparation, is crucial for predicting such terrorist activities.

\underline{Major challenges}: The following list of challenges, corresponding to functionalities/modules supported by the DTaaSS architecture, needs to be addressed for tacking such a criminal attack:

\begin{itemize}
    \item Prompt detection of suspicious UAV activity: Identifying abnormal UAV behaviors that may indicate malicious intent, particularly in the context of large public ceremonies, can provide valuable indicators for a subsequent attack.
    \item Abnormal individual human activity detection in highly-crowded places: Recognizing unusual or suspicious actions by individuals within densely populated areas can serve as already warning signs of an attack attempt.
    \item Disambiguation among numerous UAVs: Differentiating between conventional or benign UAVs and those that may be used for malicious purposes comprises a challenging and valuable line of defense.
    \item Detection of terrorism propaganda materials and signs of attack preparation on cyber channels: Monitoring online forums and social media for indications of attack planning and propaganda dissemination can provide valuable intelligence information for possible attack planning scenarios.
\end{itemize}

\section{Conclusion and future work}
\label{sec:conclusions}

In this paper, a novel Digital Twin-as-a-Security-Service (DTaaSS) architecture was introduced for holistically and significantly enhancing the protection of public spaces. The proposed framework relies on the elegant combination of a DT conceptualization with additional cutting-edge technologies, including IoT, cloud computing, Big Data analytics and AI. Specifically, DTaaSS constitutes a holistic, real-time, large-scale, comprehensive and data-driven security solution for the robust protection of public spaces, supporting: a) data collection and analytics, b) area monitoring/control and proactive threat detection, c) incident/attack prediction, and d) quantitative and data-driven vulnerability assessment. DTaaSS demonstrates increased capabilities towards addressing complex, hybrid and combined threats over large, critical and popular soft-targets. Its applicability and robustness is discussed using concrete case studies against representative and diverse real-world application scenarios, including complex attacks to: a) a metro station, b) a leisure site, and c) a cathedral square.

Regarding future research, the DTaaSS architecture can be expanded in several key aspects. One major goal comprises the implementation and evaluation of the framework using advanced simulation engines. This involves the deployment of the framework in virtual environments to emulate various security conditions and threat scenarios, providing valuable insights into improving detection algorithms, evaluating system performance and refining response processes. In parallel, addressing ethical and privacy concerns related to the collection and processing of personal data is also a priority, with a focus on implementing robust data protection measures and ensuring compliance with relevant regulations. Moreover, developing comprehensive training programs for end-users is essential for maximizing the effectiveness/usage of the DTaaSS system and for ensuring its seamless integration with existing security practices and protocols.

\balance
\bibliographystyle{IEEEtran}
\bibliography{Digital_Twins_Security}

\end{document}